\documentclass[12pt,english]{article}
\usepackage{palatino}
\usepackage[T1]{fontenc}
\usepackage[latin1]{inputenc}
\usepackage{a4wide}
\usepackage{babel}
\usepackage{graphics}

\makeatletter

\providecommand{\LyX}{L\kern-.1667em\lower.25em\hbox{Y}\kern-.125emX\@}
\let\SF@@footnote\footnote
\def\footnote{\ifx\protect\@typeset@protect
    \expandafter\SF@@footnote
  \else
    \expandafter\SF@gobble@opt
  \fi
}
\expandafter\def\csname SF@gobble@opt \endcsname{\@ifnextchar[
  \SF@gobble@twobracket
  \@gobble
}
\edef\SF@gobble@opt{\noexpand\protect
  \expandafter\noexpand\csname SF@gobble@opt \endcsname}
\def\SF@gobble@twobracket[#1]#2{}

\makeatother
\begin{document}

\title{Mathematical models of haploinsufficiency}

\author{Indrani Bose and Rajesh Karmakar}

\maketitle
{\centering Department of Physics \par}

{\centering Bose Institute\par}

{\centering 93/1, Acharya Prafulla Chandra Road, Kolkata-700 009,
India \par}

\begin{abstract}
We study simple mathematical models of gene expression to explore
the possible origins of haploinsufficiency (HI). In a diploid organism,
each gene exists in two copies and when one of these is mutated, the
amount of proteins synthesized is reduced and may fall below a threshold
level for the onset of some desired activity. This can give rise to
HI, a manifestation of which is in the form of a disease. We consider
both deterministic and stochastic models of gene expression and suggest
possible scenarios for the occurrence of HI in the two cases. In the
stochastic case, random fluctuations around the mean protein level
give rise to a finite probability that the protein level falls below
a threshold. Increased gene copy number and faster gene expression
kinetics reduce the variance around the mean protein level. The difference
between slow and fast gene expression kinetics, as regards response
to a signaling gradient, is further pointed out. The majority of results
reported in the paper are derived analytically.
\end{abstract}

\section*{I. Introduction}

Complex multicellular organisms are in general diploids, i.e., each
cell in an organism contains two copies of the full set of genes in
contrast to haploids in which each cell contains a single copy of
the genome. Genes provide the blueprint for the synthesis of proteins
which perform essential functions in cells. If one copy of a gene
is mutated, there is approximately a \( 50\% \) reduction in the
level of proteins synthesized. In many cases this does not lead to
observable changes and normalcy is retained. A common interpretation
of haploinsufficiency (HI) is that it occurs when half normal levels
of proteins are insufficient for completing particular tasks, leading
to specific types of diseases. More generally, HI may occur when the
level of proteins synthesized falls below a critical level for the
onset of some desired activity.

There is presently an extensive literature on the genetic and biomedical
aspects of HI \cite{key-61,key-62,key-63} but mathematical models
exploring the origin of HI and the issues related to it are practically
non-existent. It is by now well-accepted that stochastic processes
have considerable effect on patterns of gene expression in cells \cite{key-64,key-65,key-66,key-67}.
Cook et al \cite{key-63} have studied the role of stochastic gene
expression in HI by constructing a minimal model of gene expression
and using numerical techniques to simulate the model. Their major
finding is that when one of the two genes in a diploid organism is
inactivated due to mutations, there is an increased susceptibility
to stochastic initiations and interruptions of gene expression. As
a result, the number of proteins produced may transiently fall below
the desired level giving rise to HI. Both increased gene copy number
and faster gene expression kinetics reduce expression noise, thus
enhancing the possibility of a stable outcome.

A large number of diseases are caused by mutations in genes encoding
proteins called transcription factors (TF). More than \( 30 \) different
human maladies have been attributed to TF HI \cite{key-61}. TFs regulate
gene expression by binding at the promoter region of the gene to be
expressed. Cooperative interactions among the TFs favour the formation
of bound TF complexes (oligomers). The TFs interact at only one site
or at multiple sites of the promoter. A simple mathematical model
has been proposed to explore HI in systems involving cooperative assembly
of TFs \cite{key-62}. Such multimeric complexes are essential for
initiation of gene expression in many eukaryotic systems. The model
explores the relationship of fractional oligomerization \( Y \) with
the free (\( [S] \)) as well as total concentrations of TFs (\( [S_{0}] \)).
The TFs oligomerise to form a bound complex. The curves \( Y \) versus
\( [S] \) and \( [S_{0}] \) have sigmoidal shapes. Due to the characteristic
S shape of a sigmoid, a small change in the TF concentration around
the inflection point (the point at which the tangent to the curve
has the maximum slope) gives rise to a significant change in the magnitude
of \( Y \). Thus, if there are two TF-encoding genes and one of these
becomes silent, the level of TFs produced may fall below the inflection
point of the sigmoid and consequently the magnitude of \( Y \), the
fractional oligomerization, is considerably decreased. This results
in reduced expression from the target gene, giving rise to TF HI if
the amount of proteins synthesized falls below a threshold level. 

In Section 2 of this paper, we extend the minimal model of Cook et
al \cite{key-63} to investigate the influence of bound complexes
of TFs on the initiation of gene expression. In Section 3, we study
the stochastic version of the minimal model and its extensions to
elucidate the role of stochasticity in HI. We derive analytical expressions
for the quantities determined numerically by Cook et al \cite{key-63}.

\section*{2. Deterministic Model}

The model is an extension of the minimal model of gene expression
studied by Cook et al \cite{key-63}. A brief description of the model
is as follows. A gene can be in two possible states: inactive (\( G \))
and active (\( G^{*} \)). Random transtions occur between the states
\( G \) and \( G^{*} \) according to the first order reaction kinetics\begin{equation}
\label{mathed:first-eqn}
G\quad \begin{array}{c}
k_{a}\\
\rightleftharpoons \\
k_{d}
\end{array}\quad G^{\star }\quad \begin{array}[b]{c}
j_{p}\\
\longrightarrow 
\end{array}\quad p\quad \begin{array}[b]{c}
k_{p}\\
\longrightarrow 
\end{array}\quad \Phi 
\end{equation}

{\raggedright where \( k_{a} \) and \( k_{d} \) are the activation
and deactivation rate constants. The corresponding half-times are
\( T_{a}=\frac{log2}{k_{a}} \) and \( T_{d}=\frac{log2}{k_{d}} \)
respectively. In the active state \( G^{*} \), the gene synthesizes
a protein (\( p \)) with the rate constant \( j_{p} \). The protein
product degrades with a rate constant \( k_{p} \) and the associated
half-time is \( T_{p} \). The protein degradation product is represented
as \( \Phi  \). We now assume that activation to the state \( G^{*} \)
is brought about by an inducing stimulus \( S \), e.g., TFs. The
reaction scheme in the presence of the stimulus is given by \begin{equation}
\label{mathed:second-eqn}
G+S\quad \begin{array}{c}
k_{1}\\
\rightleftharpoons \\
k_{2}
\end{array}\quad GS\quad \begin{array}{c}
k_{a}\\
\rightleftharpoons \\
k_{d}
\end{array}\quad G^{\star }\quad \begin{array}[b]{c}
j_{p}\\
\longrightarrow 
\end{array}\quad p\quad \begin{array}[b]{c}
k_{p}\\
\longrightarrow 
\end{array}\quad \Phi 
\end{equation}
\par}

{\raggedright where \( GS \) represents the bound complex of \( G \)
and \( S \) from which transition to the active state \( G^{*} \)
occurs. If \( n_{G} \) is the total concentration of genes then\begin{equation}
\label{mathed:third-eqn}
n_{G}=[G]+[GS]+[G^{*}]
\end{equation}
\par}

{\raggedright where \( [G] \), \( [GS] \) and \( [G^{*}] \) denote
the concentrations of genes in the states \( G \), \( GS \) and
\( G^{*} \) respectively. In the steady state, we have \[
\frac{d[G]}{dt}=k_{2}[GS]-k_{1}[G][S]=0\]
\par}

so that \begin{equation}
\label{mathed:fourth-eqn}
\frac{[G][S]}{K_{s}}=[GS]
\end{equation}

{\raggedright where \( K_{s}=\frac{k_{2}}{k_{1}} \) is the equilibrium
dissociation constant. From (\ref{mathed:third-eqn}) and (\ref{mathed:fourth-eqn}),
we get \begin{equation}
\label{mathed:fifth-eqn}
[GS]=\frac{n_{G}[S]/K_{s}}{1+[S]/K_{s}}-[G^{*}]\frac{[S]/K_{s}}{1+[S]/K_{s}}
\end{equation}
\par}

{\raggedright Also, in the steady state, \begin{equation}
\label{mathed:sixth-eqn}
\frac{d[G^{*}]}{dt}=k_{a}[GS]-k_{d}[G^{*}]=0
\end{equation}
\par}

{\raggedright From (\ref{mathed:fifth-eqn}) and (\ref{mathed:sixth-eqn}),
the expression for \( [G^{*}] \) in the steady state is given by
\begin{equation}
\label{mathed:seventh-eqn}
[G^{*}]=\frac{n_{G}k_{a}\frac{[S]/K_{s}}{1+[S]/K_{s}}}{k_{a}\frac{[S]/K_{s}}{1+[S]/K_{s}}+k_{d}}
\end{equation}
\par}

{\raggedright The reaction scheme in (\ref{mathed:first-eqn}) leads
to the expression\begin{equation}
\label{mathed:eighth-eqn}
[G^{*}]=\frac{n_{G}k_{a}}{k_{a}+k_{d}}
\end{equation}
\par}

{\raggedright in the steady state. Expression (\ref{mathed:seventh-eqn})
and (\ref{mathed:eighth-eqn}) are equivalent on defining effective
activation and deactivation rate constants:\[
k_{a}^{'}=k_{a}\frac{[S]/K_{s}}{1+[S]/K_{s}},\]
\begin{equation}
\label{mathed:nineth-eqn}
k_{d}^{'}=k_{d}
\end{equation}
 We now assume the inducing stimulus to be TFs. In the simplest approximation,
\( n \) individual TFs oligomerise to produce an active complex \( S_{n} \)
according to the reaction scheme \begin{equation}
\label{mathed:tenth-eqn}
n\, S\quad \begin{array}[b]{c}
K\\
\rightleftharpoons 
\end{array}\quad S_{n}
\end{equation}
\par}

{\raggedright The \( n \) TFs interact all at once to give rise to
the bound complex \( [S_{n}] \), i.e., we ignore the formation of
dimers, tetramers, ..... etc. Let \( [S_{0}] \) be the initial concentration
of TFs. Then \begin{equation}
\label{mathed:eleventh-eqn}
[S_{0}]=[S]+n\, [S_{n}]
\end{equation}
\par}

{\raggedright where \( [S] \) and \( [S_{n}] \) are the concentrations
of free TFs and the bound TF-complex respectively. The global equilibrium
constant \( K \) is given by\begin{equation}
\label{mathed:twelveth-eqn}
K=\frac{[S_{n}]}{[S]^{n}}=\frac{[S_{n}]}{([S_{0}]-n\, [S_{n}])^{n}}
\end{equation}
\par}

{\raggedright The fractional oligomerization is defined as \cite{key-62}
\begin{equation}
\label{mathed:thirteenth-eqn}
Y=\frac{n\, [S_{n}]}{[S_{0}]}=\frac{n\, [S_{n}]}{[S]+n\, [S_{n}]}
\end{equation}
\par}

{\raggedright Using (\ref{mathed:eleventh-eqn}) and (\ref{mathed:twelveth-eqn}),
\( Y \) can further be written as\begin{equation}
\label{mathed:14-eqn}
Y=\frac{[S]^{n}}{\frac{[S]}{K\, n}+[S]^{n}}
\end{equation}
\par}

and \begin{equation}
\label{mathed:eqn-15}
Y=\frac{([S_{0}]-n\, [S_{n}])^{n}}{\frac{([S_{0}]-n\, [S_{n}])}{K\, n}+([S_{0}]-n\, [S_{n}])^{n}}
\end{equation}

{\raggedright Fig.1 shows the curve, fractional oligomerization \( Y \)
versus \( [S_{0}] \) ( Eq. (\ref{mathed:eqn-15}) ) for \( n=6,K=2 \)
and \( [S_{n}]=0.2 \). The curve has the well-known sigmoidal shape. \par}

\begin{figure}
{\centering \resizebox*{3in}{!}{\includegraphics{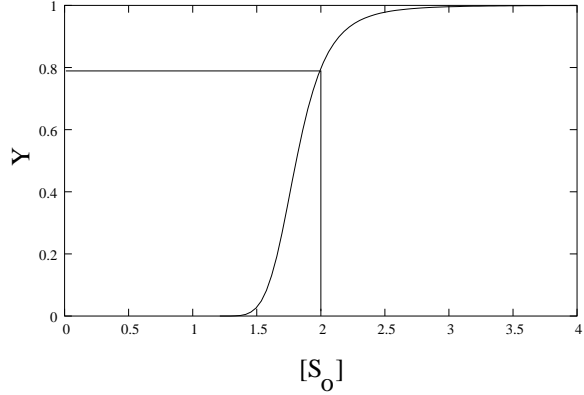}} \par}

\caption{Fractional oligomerization \protect\( Y\protect \) versus \protect\( [S_{0}]\protect \)
(Eq. (\ref{mathed:eqn-15})) for \protect\( n=6\protect \), \protect\( K=2\protect \)
and}

\( [S_{n}]=0.2 \)
\end{figure}

We now replace \( [S] \) by \( [S_{n}] \) in the reaction scheme
described by Eq. (\ref{mathed:second-eqn}), i.e., we assume that
the TF-oligomer \( S_{n} \) binds to a gene in the inactive state
\( G \) to give rise to the bound complex \( GS_{n} \). Transition
to the active state \( G^{*} \) occurs from the intermediate state
\( GS_{n} \). The concentration \( [G^{*}] \) in the steady state
is obtained from (\ref{mathed:seventh-eqn}) by replacing \( [S] \)
by \( [S_{n}] \) where \( [S_{n}]=K\, [S]^{n} \) ( Eq. (\ref{mathed:twelveth-eqn})
). One finally obtains \begin{equation}
\label{mathed:eqn-16}
[G^{*}]=\frac{n_{G}\, k_{a}\, \frac{K\, [S]^{n}/K_{s}}{1+K\, [S]^{n}/K_{s}}}{k_{a}\, \frac{K\, [S]^{n}/K_{s}}{1+K\, [S]^{n}/K_{s}}+k_{d}}
\end{equation}

{\raggedright The concentration of proteins in the steady state is
given by\begin{equation}
\label{mathed:eqn-17}
[p]=\frac{j_{p}}{k_{p}}\, [G^{*}]
\end{equation}
\par}

{\raggedright From (\ref{mathed:14-eqn}), \( [S]^{n} \) can be written
as \begin{equation}
\label{mathed:eqn-18}
[S]^{n}=\frac{Y}{1-Y}\, \frac{[S]}{K\, n}
\end{equation}
\par}

{\raggedright From (\ref{mathed:eqn-16}) and (\ref{mathed:eqn-17})
and for \( k_{a}=k_{d} \), we get\begin{equation}
\label{mathed:eqn-19}
[p]=n_{G}\, \frac{j_{p}}{k_{p}}\, \frac{a\, Y}{1+2aY-Y}
\end{equation}
\par}

where \( a=\frac{[S]}{n\, K_{s}}. \)

{\raggedright Fig. 2 shows the protein concentration \( [p] \) versus
fractional oligomerization \( Y \) for \( k_{a}=k_{d} \) (Eq.(\ref{mathed:eqn-19})),
\( n_{G}=2 \), \( j_{p}=0.5\, k_{p} \), \( K_{s}=1.2 \) and \( n=6 \). 
\begin{figure}
{\centering \resizebox*{3in}{!}{\includegraphics{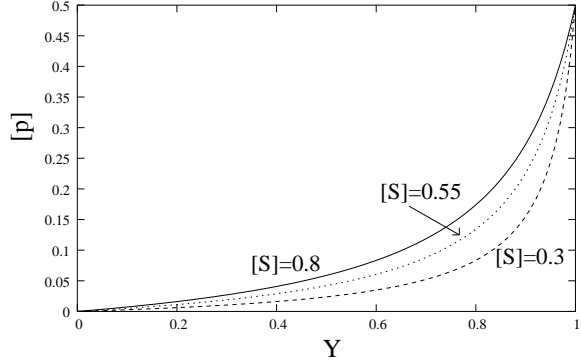}} \par}

\caption{Protein concentration \protect\( [p]\protect \) versus fractional
oligomerization \protect\( Y\protect \) for \protect\( k_{a}=k_{d}\protect \)
(Eq. (\ref{mathed:eqn-19})), \protect\( n_{G}=2\protect \), \protect\( j_{p}=0.5\, k_{p}\protect \),
\protect\( K_{s}=1.2\protect \) and \protect\( n=6\protect \). }
\end{figure}
 From Eqs. (\ref{mathed:eleventh-eqn}), (\ref{mathed:eqn-16}), and
(\ref{mathed:eqn-17}) and for \( k_{a}=k_{d} \), the concentration
of protein \( [p] \), as a function of the total concentration \( [S_{0}] \)
of TFs, can be written as\begin{equation}
\label{mathed:eqn-20}
[p]=n_{G}\, \frac{j_{p}}{k_{p}}\, \frac{([S_{0}]-n\, [S_{n}])^{n}}{2\, ([S_{0}]-n\, [S_{n}])^{n}+K_{s}/K}
\end{equation}
\par}

\begin{figure}
{\centering \resizebox*{3in}{!}{\includegraphics{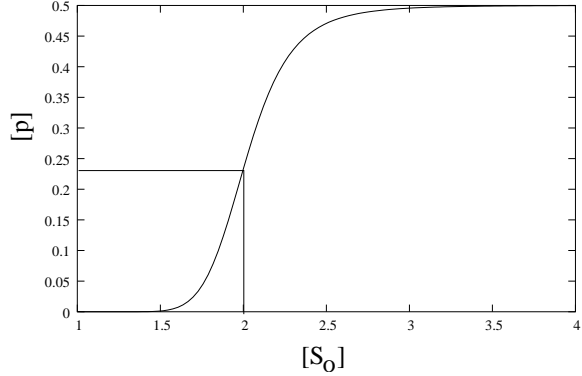}} \par}

\caption{Protein concentration \protect\( [p]\protect \) versus \protect\( [S_{0}]\protect \)
(Eq. (\ref{mathed:eqn-20}))for \protect\( n=6\protect \), \protect\( K=2\protect \),
\protect\( K_{s}=1.2\protect \) and \protect\( [S_{n}]=0.2\protect \)}
\end{figure}

{\raggedright Fig. 3 shows the plot \( [p] \) versus \( [S_{0}] \)
(Eq. (\ref{mathed:eqn-20})) for \( n=6 \), \( K=2 \), \( K_{s}=1.2 \)
and \( [S_{n}]=0.2 \). Figs. 1, 2 and 3 provide a possible explanation
for the origin of HI. Suppose two TF-encoding genes produce TFs of
total concentration \( [S_{0}]=4 \). If one of the genes is inactivated
due to mutations, the total concentration \( [S_{0}] \) falls to
the value \( 2 \). For the parameter values corresponding to Fig.
1, the fractional oligomerization \( Y \) has the value \( 0.797 \).
The TFs form a bound complex \( S_{n} \) (\( n=6 \)) which then
activates the gene synthesizing the protein \( P \). The concentration
of proteins \( [p] \) corresponding to \( Y=0.797 \) and for \( [S]=0.8 \)
is given by \( [p]=0.233 \) (Fig. 2). The same value \( [p] \) is
obtained from Fig. 3 with \( [S_{0}]=2. \) If both the encoding genes
are active, the values of \( [S_{0}] \), \( Y \) and \( P \) are
\( [S_{0}]=4, \) \( Y=1.0 \) and \( [p]=0.5 \) respectively. Thus,
for one gene, the protein level is reduced by more than half. If the
level falls below a threshold, the amount of proteins synthesized
is not sufficient for the execution of a particular task. This gives
rise to HI, the manifestation of which is in the form of a disease.\par}

\section*{3. Stochastic Approach}

Let us first consider the model described in Section 2 in the absence
of an inducing stimulus. Let \( n_{tot} \) be the total no of genes
and \( n_{0} \) , \( n_{1} \) (\( n_{tot}=n_{0}+n_{1} \)), the
number of genes in the inactive (\( G \)) and active (\( G^{*} \))
states respectively. In the stochastic model, a gene makes random
transitions between the inactive and active states with \( k_{a} \)
and \( k_{d} \) being the activation and deactivation rate constants.
In the active state, protein production and degradation occur with
the rate constants \( j_{p} \) and \( k_{p} \) respectively. Let
\( p(n_{1},n_{2},t) \) be the probability that at time \( t, \)
\( n_{1} \) genes are in the active state \( G^{*} \) and the number
of protein molecules is \( n_{2}. \) The rate of change of the probability
with respect to time is given by the Master Equation

\begin{equation}
\label{mathed:eqn-21}
\begin{array}{cc}
\frac{\partial p(n_{1},n_{2},t)}{\partial t}= & k_{a}[(n_{tot}-n_{1}+1)p(n_{1}-1,n_{2},t)-(n_{tot}-n_{1})p(n_{1},n_{2},t)]\\
 & +k_{d}[(n_{1}+1)p(n_{1}+1,n_{2},t)-n_{1}p(n_{1},n_{2},t)]\\
 & +j_{p}[n_{1}p(n_{1},n_{2}-1,t)-n_{1}p(n_{1},n_{2},t)]\\
 & +k_{p}[(n_{2}+1)p(n_{1},n_{2}+1,t)-n_{2}p(n_{1},n_{2},t)]
\end{array}
\end{equation}

{\raggedright For each rate constant, there is a gain term which adds
to the probability and a loss term which subtracts from the probability. \par}

We now use the standard approach in the theory of stochastic processes
\cite{key-68} to determine the average number of activated genes
\( <n_{1}> \) and proteins \( <n_{2}> \) in the steady state and
the variances thereof. Define the generating function \begin{equation}
\label{mathed:eqn-22}
F(z_{1},\, z_{2},\, t)=\sum _{n_{1},n_{2}}z_{1}^{n_{1}}\, z_{2}^{n_{2}}\, p(n_{1},\, n_{2},t)
\end{equation}
In terms of the generating function, the Master equation (\ref{mathed:eqn-21})
becomes\begin{equation}
\label{mathed:eqn-23}
\frac{\partial F}{\partial t}=k_{a}n_{tot}(z_{1}-1)F-k_{a}(z_{1}-1)z_{1}\frac{\partial F}{\partial z_{1}}-k_{d}(z_{1}-1)\frac{\partial F}{\partial z_{1}}+j_{p}(z_{2}-1)z_{1}\frac{\partial F}{\partial z_{1}}-k_{p}(z_{2}-1)\frac{\partial F}{\partial z_{2}}
\end{equation}

{\raggedright In the steady state \( \frac{\partial F}{\partial t}=0. \)
The following properties of the generating function are used in subsequent
calculations:\begin{equation}
\label{mathed:eqn-24}
F\mid _{z_{1}=1,z_{2}=1}=1
\end{equation}
\begin{equation}
\label{mathed:eqn-25}
<n_{1}>=\frac{\partial F}{\partial z_{1}}\mid _{z_{1}=1,z_{2}=1},\: \: \: <n_{2}>=\frac{\partial F}{\partial z_{2}}\mid _{z_{1}=1,z_{2}=1}
\end{equation}
where \( <n_{1}> \) is the mean number of active genes, i.e., genes
in the state \( G^{*} \) and \( <n_{2}> \) is the same for proteins.
Furthermore, \begin{equation}
\label{mathed:eqn-26}
\begin{array}{c}
\frac{\partial ^{2}F}{\partial z_{1}^{2}}\mid _{z_{1}=1,z_{2}=1}=<n_{1}^{2}>-<n_{1}>\\
\frac{\partial ^{2}F}{\partial z_{2}^{2}}\mid _{z_{1}=1,z_{2}=1}=<n_{2}^{2}>-<n_{2}>
\end{array}
\end{equation}
Hence the variances around the mean levels are given by \begin{equation}
\label{mathed:eqn-27}
\begin{array}{c}
Var_{n_{1}}=<n_{1}^{2}>-<n_{1}>^{2}=\frac{\partial ^{2}F}{\partial z_{1}^{2}}\mid _{z_{1}=1,z_{2}=1}+<n_{1}>-<n_{1}>^{2}\\
Var_{n_{2}}=<n_{2}^{2}>-<n_{2}>^{2}=\frac{\partial ^{2}F}{\partial z_{2}^{2}}\mid _{z_{1}=1,z_{2}=1}+<n_{2}>-<n_{2}>^{2}
\end{array}
\end{equation}
Successive differentiation of Eq. (\ref{mathed:eqn-23}) (\( \frac{\partial F}{\partial t}=0. \))
with respect to \( z_{1} \) and \( z_{2} \) gives rise to linear
equations for successively higher moments. The equations may be solved
to obtain, in particular, the mean and the variance. For example,
differentiating Eq. (\ref{mathed:eqn-23}) with respect to \( z_{1} \)
and \( z_{2} \) and then putting \( z_{1},\, z_{2}=1 \), one obtains
expressions for the mean.\par}

{\raggedright The mean and variance are given by \begin{equation}
\label{mathed:eqn-28}
<n_{1}>=\frac{n_{tot}\, k_{a}}{k_{a}+k_{d}}
\end{equation}
\begin{equation}
\label{mathed:eqn-29}
Var_{n_{1}}=<n_{1}>\, \frac{k_{d}}{k_{a}+k_{d}}
\end{equation}
\begin{equation}
\label{mathed:eqn-30}
<n_{2}>=<p>=<n_{1}>\, \frac{j_{p}}{k_{p}}=\frac{j_{p}}{k_{p}}\, \frac{n_{tot}\, k_{a}}{k_{a}+k_{d}}
\end{equation}
\begin{equation}
\label{mathed:eqn-31}
Var_{n_{2}}=<n_{1}>\, \frac{j_{p}}{k_{p}}\, [1+\frac{j_{p}\, k_{d}}{(k_{a}+k_{d})(k_{a}+k_{d}+k_{p})}]
\end{equation}
\par}

{\raggedright As in Ref. 3, temporal quantities are scaled relative
to the product half-life \( T_{p}=\frac{log2}{k_{p}}. \) Let \( T_{a}=\frac{log2}{k_{a}} \)
and \( T_{d}=\frac{log2}{k_{d}} \) be the times for half-maximal
gene activation and deactivation respectively. The times \( T_{a} \)
and \( T_{d} \) are scaled relative to \( T_{p} \). Some of the
results obtained in Ref. 3, using numerical simulation techniques,
can readily be derived from the analytical expressions in (\ref{mathed:eqn-28})-(\ref{mathed:eqn-31}).
Stochasticity introduces random fluctuations around the mean protein
level and variance gives a measure of the spread. Let \( T_{a}=T_{d}=T_{p}/4 \),
i.e., \( k_{a}=k_{d}=\alpha \, k_{p} \) with \( \alpha >0 \). As
\( \alpha  \) increases, one has faster expression kinetics and from
(\ref{mathed:eqn-31}) it is easy to verify that variance is reduced,
i.e., the expression noise is less. The mean product level (Eq. (\ref{mathed:eqn-30}))
is, however, independent of \( \alpha  \). With increase in \( j_{p} \),
i.e., the protein synthesis rate, the variance increases. Let us now
consider the case when the net expression rate of \( n_{tot} \) genes
is distributed to one single gene so that the mean protein level remains
the same. From (\ref{mathed:eqn-30}) \begin{equation}
\label{mathed:eqn-32}
<n_{2}>=\frac{j_{p}}{k_{p}}\, \frac{n_{tot}}{2}=\frac{j_{p}^{'}}{k_{p}}\, \frac{1}{2}
\end{equation}
\par}

{\raggedright where \( j_{p}^{'}=j_{p}\, n_{tot} \) is the expression
rate when only one gene is considered. Since \( j_{p}^{'}>j_{p} \),
the gene copy number is reduced from \( n_{tot} \) to \( 1 \). Similarly,
when the net expression rate of \( n_{tot} \) genes is distributed
to a larger number genes, say, from two to four, the variance is reduced.When
one of two genes is inactivated due to mutations, the average protein
level in the steady state is reduced by \( 50\% \). This may still
be higher than the threshold level required for protein activity.
Due to the variance around the mean level, the number of proteins
may transiently fall below the threshold giving rise to HI. The occurrence
of HI further becomes more probable for slower expression kinetics
as then the variance is increased in magnitude. For stochastic gene
expression in the presence of an inducing stimulus, say, TF's, we
use the effective model with the activation/deactivation rate constants
given in Eq. (\ref{mathed:nineth-eqn}). The expressions for the mean
and the variance are the same as in Eqs. (\ref{mathed:eqn-28})-(\ref{mathed:eqn-31})
but with \( k_{a} \), \( k_{d} \) replaced by \( k_{a}^{'} \) and
\( k_{d}^{'} \) respectively.\par}

We now derive expressions for the probability distributions of protein
levels in the steady state. To do this, we consider a simpler stochastic
model in which the only stochasticity arises from the random transitions
of a gene between the inactive and active states. In each state of
the gene, the concentration of proteins evolves deterministically
according to the equation\begin{equation}
\label{mathed:eqn-33}
\frac{dx}{dt}=\frac{j_{p}}{X_{max}}\, z-k_{p}x=f(x,z)
\end{equation}

{\raggedright where \( z=1 \) \( (0) \) when the gene is in the
active (inactive) state and \( x=\frac{X}{X_{max}}, \) \( X \) and
\( X_{max} \) being the protein concentration at time t and the maximum
protein concentration respectively. We note that \( X_{max}=\frac{j_{p}}{k_{p}}. \)
Let \( p_{j}(x,t) \) \( (j=0,1) \) be the probability density function
when \( z=j. \) The total probability density function is \begin{equation}
\label{mathed:eqn-34}
p(x,t)=p_{0}(x,t)+p_{1}(x,t)
\end{equation}
\par}

{\raggedright The rate of change of probability density is given by\begin{equation}
\label{mathed:eqn-35}
\frac{\partial p_{j}(x,t)}{\partial t}=-\frac{\partial }{\partial x}[f(x,j)p_{j}(x,t)]+\sum _{k\neq j}[W_{kj}\, p_{k}(x,t)-W_{jk}\, p_{j}(x,t)]
\end{equation}
where \( W_{kj} \) is the transition rate from the state \( k \)
to the state \( j \) and \( W_{jk} \) is the same for the reverse
transition. The first term in Eq. (\ref{mathed:eqn-35}) is the so
called {}``transport'' term representing the net flow of the probability
density. The second term represents the gain/loss in the probability
density due to random transitions between the state \( j \) and other
accessible states. In the present case, Eq. (\ref{mathed:eqn-35})
gives rise to the following two equations: \begin{equation}
\label{mathed:eqn-36}
\frac{\partial p_{0}(x,t)}{\partial t}=-\frac{\partial }{\partial x}(-k_{p}x\, p_{0}(x,t))+k_{d}\, p_{1}(x,t)-k_{a}\, p_{0}(x,t)
\end{equation}
\begin{equation}
\label{mathed:eqn-37}
\frac{\partial p_{1}(x,t)}{\partial t}=-\frac{\partial }{\partial x}\{(\frac{j_{p}}{X_{max}}-k_{p}\, x)p_{1}(x,t)\}+k_{a}\, p_{0}(x,t)-k_{d}\, p_{1}(x,t)
\end{equation}
The Master equation (Eq. (\ref{mathed:eqn-21})) provides a full stochastic
description of all the processes associated with gene expression,
namely, gene activation and deactivation, protein synthesis and degradation.
The only stochastic events considered by Cook et al \cite{key-63}
are those related to gene activation and deactivation. In their model,
protein synthesis from the active gene and protein degradation occur
in a deterministic manner. Eqs. (\ref{mathed:eqn-36}) - (\ref{mathed:eqn-37})
describe the scenario studied by Cook et al. The steady state distribution
in this case is given by \begin{equation}
\label{mathed:eqn-38}
p(x)=C\, x^{(\frac{k_{a}}{k_{p}}-1)}(1-x)^{(\frac{k_{d}}{k_{p}}-1)}
\end{equation}
where \( C \) the normalization constant is given by the inverse
of a beta function. \begin{equation}
\label{mathed:eqn-39}
C=\frac{1}{B(\frac{k_{a}}{k_{p}},\frac{k_{d}}{k_{p}})}
\end{equation}
\par}

\begin{figure}
{\centering \resizebox*{3in}{!}{\includegraphics{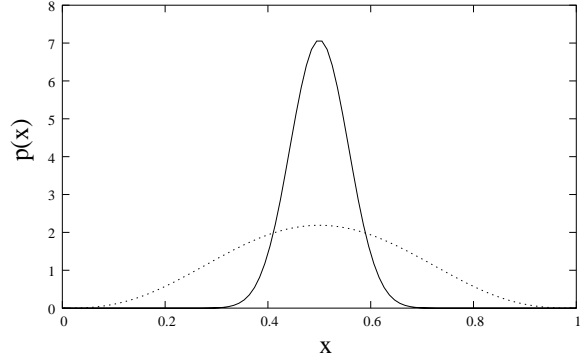}} \par}

\caption{\protect\( p(x)\protect \) versus \protect\( x\protect \) for slow
(dotted curve) and fast (solid) gene expression kinetics}
\end{figure}

{\raggedright Since the probability density function is known, the
mean protein level and its variance can be calculated in a straightforward
manner. The mean protein level is identical to that obtained from
the Master equation (Eq. (\ref{mathed:eqn-21})) whereas the variance
is underestimated as stochasticity is taken into account only at the
levels of gene activation and deactivation. Fig. 4 shows the plot
of \( p(x) \) versus \( x \) for slow (\( T_{a}=T_{d}=T_{p}/4 \))
and fast (\( T_{a}=T_{d}=T_{p}/40 \)) gene expression kinetics. In
the latter case, the distribution is significantly narrower, i.e.,
faster kinetics lead to a reduction in the variance. The same conclusion
is reached from the Master equation approach. From the full width
at half maximum of the broader distribution, one finds that \( x \)
ranges from 0.25 to 0.75, 0.5 being the mean value. Thus, it is probable
that the protein level falls below the threshold for desired activity
giving rise to HI.  Let \( x_{thr} \) (\( <1 \)) be the threshold
value of \( x \). The probability that \( x \) is greater than \( x_{thr} \)
is \begin{equation}
\label{mathed:eqn-40}
p(x>x_{thr})=1-\frac{\int _{0}^{x_{thr}}x^{(\frac{k_{a}}{k_{p}}-1)}(1-x)^{(\frac{k_{d}}{k_{p}}-1)}dx}{\int _{0}^{1}x^{(\frac{k_{a}}{k_{p}}-1)}(1-x)^{(\frac{k_{d}}{k_{p}}-1)}dx}
\end{equation}
\begin{equation}
\label{mathed:eqn-41}
=1-\frac{_{k_{p}\, x_{thr}^{\frac{k_{a}}{k_{p}}}\, _{2}F_{1}[1-\frac{k_{d}}{k_{p}},\frac{k_{a}}{k_{p}},1+\frac{k_{a}}{k_{p}},X_{thr}]}}{k_{a}\, B(\frac{k_{a}}{k_{p}},\frac{k_{d}}{k_{p}})}
\end{equation}
where \( _{2}F_{1}(a,b,c;z) \) is the hypergeometric function.\par}

{\raggedright Let \( x_{thr} \) be \( 0.25 \). The probability \( p(x>x_{thr}) \)
is computed using Mathematica for both slow and fast gene expression
kinetics. The values of \( p(x>x_{thr}) \) in the slow and fast cases
are \( 0.9294 \) and \( 0.9999 \) respectively. Since in the latter
case, the probability that the protein level exceeds the threshold
is higher, the chance of HI occurrence is correspondingly lower.\par}

\begin{figure}
{\centering \resizebox*{3in}{!}{\includegraphics{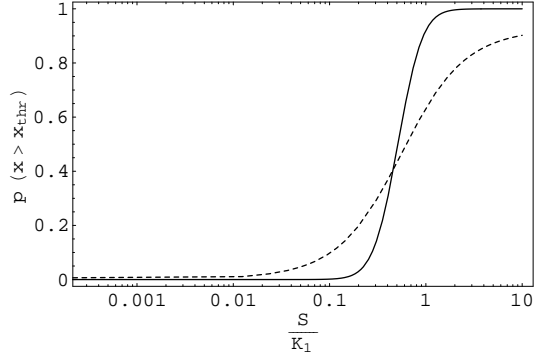}} \par}

\caption{\protect\( p(x>x_{thr})\protect \) versus \protect\( S/K_{s}\protect \)
in a semi-logarithm plot for \protect\( n=1.\protect \) The solid
(dotted) curve corresponds to fast (slow) kinetics.}
\end{figure}

\begin{figure}
{\centering \resizebox*{3in}{!}{\includegraphics{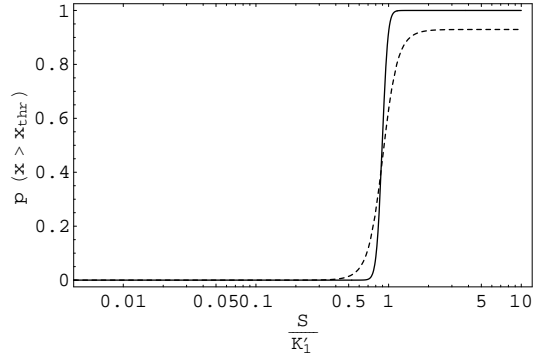}} \par}

\caption{\protect\( p(x>x_{thr})\protect \) versus \protect\( S/K^{'}_{s}\protect \)
{[}\protect\( (K_{s}^{'})^{n}=K_{s}/K\protect \){]} in a semi-logarithm
plot for \protect\( n=6.\protect \) The solid (dotted) curve corresponds
to fast (slow) kinetics.}
\end{figure}

In the presence of an inducing stimulus, say, TFs, the probability
of activation above a threshold is again given by (\ref{mathed:eqn-40})
with \( k_{a} \) and \( k_{d} \) replaced by \( k_{a}^{'} \) and
\( k_{d}^{'} \) (Eq. (\ref{mathed:nineth-eqn})) . Fig. 5 shows \( p(x>x_{thr}), \)
\( x_{thr}=0.25 \), versus \( S/K_{s} \) in a semi-logarithm plot
for both slow and fast kinetics. In the fast case, a substantially
steeper curve is obtained leading to enhanced signal discrimination,
i.e., a more predictable response in a gradient of inducing signal.
 As shown by Cook et al \cite{key-63}, the signal discrimination
ability increases with gene copy number. One can thus speculate that
diploid organisms utilise stochastic expression kinetics, preferably
fast, for signal discrimination and are susceptible to degraded signal
discrimination due to a reduction of gene copy number in the haploid
state. Mutations in the subset of genes which generate a response
to signaling gradients in diploid organisms may be the cause of some
HI syndromes associated with these systems. Fig. 6 shows the same
plot as in Fig. 5 but now the TF's form bound complexes with \( n=6. \)
In Eq. (\ref{mathed:nineth-eqn}), \( [S] \) is replaced by \( [S_{n}]=K[S]^{n} \)
( Eq. (\ref{mathed:twelveth-eqn})). One now finds that the slopes
of the curves for slow and fast kinetics are similar. Thus as \( n \),
the number of TF's forming the bound complex, increases, the distinction
between slow and fast gene expression kinetics, as regards their signal
discrimination ability, becomes less pronounced.

\section*{4. Summary}

In this paper, we have studied simple mathematical models to explore
the possible origins of HI. In Section 2, we have considered a deterministic
model in which a complex of \( n \) TFs binds at the appropriate
region of DNA to initiate gene expression in eukaryotes. The concentration
of proteins synthesized versus the total concentration of TFs (Fig.
3) is a sigmoid. Due to the S-shape of the curve, the protein level
may fall below a threshold when one of the two genes synthesizing
the TFs is mutated, resulting in a \( 50\% \) reduction in the total
concentration of TFs. The absence of required protein activity can
give rise to HI. In Section 3, we have studied simple stochastic models
of gene expression and shown that due to random fluctuations around
the mean protein concentration in the steady state, the protein level
may fall below the threshold even though it does not do so in the
deterministic case. The variance, a measure of the spread around the
mean protein level, is reduced with increasing gene copy number and
faster expression kinetics. The variance increases if the rate constant
\( j_{p} \) associated with protein synthesis is increased. In the
case of one gene, we have further calculated the probability that
the concentration of proteins exceeds a threshold in the absence as
well as the presence of an inducing stimulus. In the latter case,
faster gene expression kinetics give rise to a sharper response to
changing stimulus concentrations. As shown by Cook et al \cite{key-63},
this is also true when the gene copy number is increased. Thus the
signal discrimination ability of diploid organisms may be impaired
in the haploid state. When the inducing stimulus is a bound complex
of \( n \) TFs, the distinction between slow and fast gene expression
kinetics becomes less with increasing \( n \). To sum up, we have
considered both deterministic as well as stochastic models of gene
expression and indicated possible scenarios for the occurrence of
HI.

{\centering \textbf{ACKNOWLEDGEMENT}\par}

R. K. was supported by the Council of Scientific and Industrial Research,
India under Sanction No. 9/15 (239) / 2002 - EMR - 1.

\end{document}